\RequirePackage{fix-cm}
\documentclass[twocolumn]{svjour3}          
\smartqed  
\usepackage{amsmath}
\usepackage{graphicx}
\usepackage{xcolor}
\usepackage{verbatim}
\usepackage{hyperref}
\usepackage{fancyhdr}

\fancypagestyle{specialfooter}{%
  \fancyhf{}
  
  \fancyfoot[C]{\bf\emph{Published in Computer Graphics International, $12^{th}$-$15^{th}$ June 2012, Bournemouth, UK}}
}

\begin{document}
\cfoot{\bf\emph{Published in Computer Graphics International, $12^{th}$-$15^{th}$ June 2012, Bournemouth, UK}}

\title{A learning-based approach for automatic image and video colorization}
\subtitle{}

\author{Raj Kumar Gupta \and Alex Yong-Sang Chia \and Deepu Rajan \and Huang Zhiyong}
\institute{Raj Kumar Gupta \and Deepu Rajan \at School of Computer Engineering, Nanyang Technological University - Blk N4, Nanyang Avenue Singapore 639798 \\E-mail: \{ rajk0005, asdrajan \}@ntu.edu.sg  \and Alex Yong-Sang Chia \and Huang Zhiyong \at Institute for Infocomm Research, 1 Fusionopolis Way, \#21-01 Connexis (South Tower), Singapore 138632 \\E-mail: \{ ysachia, zyhuang \}@i2r.a-star.edu.sg}
\date{ }

\maketitle

\begin{abstract}
In this paper, we present a color transfer algorithm to colorize a broad range of gray images without any user intervention. The algorithm uses a machine learning-based approach to automatically colorize grayscale images. The algorithm uses the superpixel representation of the reference color images to learn the relationship between different image features and their corresponding color values. We use this learned information to predict the color value of each grayscale image superpixel. As compared to processing individual image pixels, our use of superpixels helps us to achieve a much higher degree of spatial consistency as well as speeds up the colorization process. The predicted color values of the gray-scale image superpixels are used to provide a \emph{'micro-scribble'} at the centroid of the superpixels. These color scribbles are refined by using a voting based approach. To generate the final colorization result, we use an optimization-based approach to smoothly spread the color scribble across all pixels within a superpixel. Experimental results on a broad range of images and the comparison with existing state-of-the-art colorization methods demonstrate the greater effectiveness of the proposed algorithm.
 
\end{abstract}

\begin{figure}
\centering
\begin{tabular}{c@{\hspace{1.05mm}} c@{\hspace{0mm}}}
			\includegraphics[width=0.48\columnwidth]{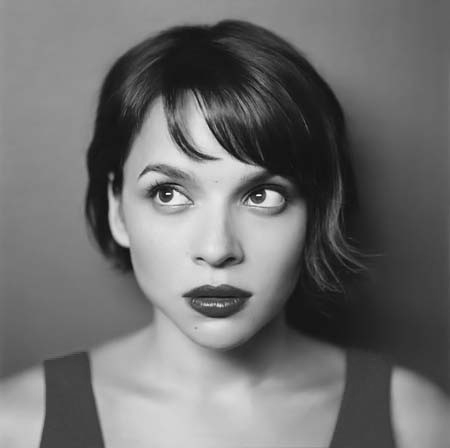}& 
			\includegraphics[width=0.48\columnwidth]{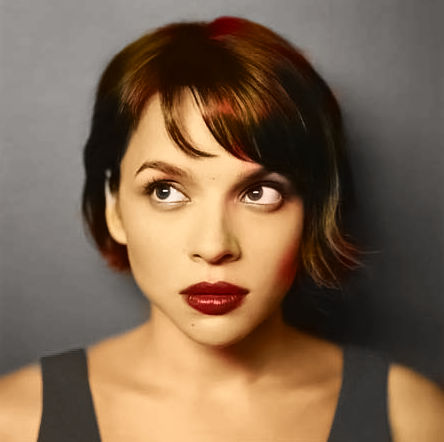}\\
			\includegraphics[width=0.48\columnwidth]{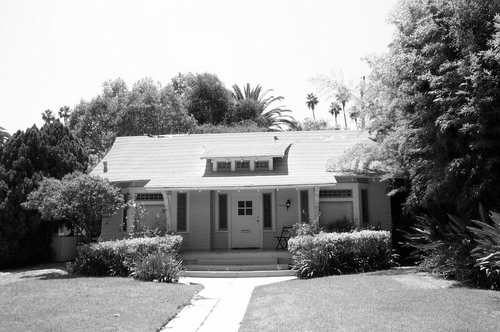}& 
			\includegraphics[width=0.48\columnwidth]{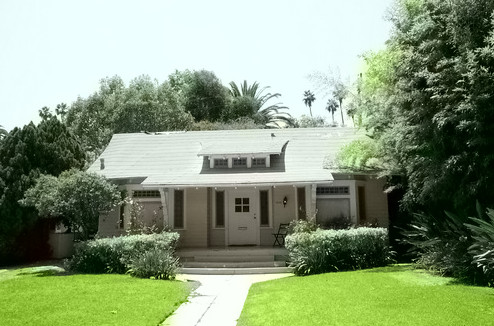}\\					
\end{tabular}
\caption{Colorization results generated by using the proposed algorithm without any user intervention. The first column shows the grayscale images and the second column shows the colorization results.}
\label{fig:1}
\end{figure}

\thispagestyle{specialfooter}

\section{Introduction}
\label{sec:1}
Image colorization is a process that adds color to black-and-white images, requiring an extensive amount of user intervention. Traditionally, the colorization process is very tedious, time consuming and requires $prior$ knowledge and artistic skills to assign the appropriate colors to the grayscale image. Recently, several colorization algorithms~\cite{Welsh02}~\cite{Levin04}~\cite{Irony05}~\cite{Yatziv06}~\cite{Szeliski06}~\cite{Luan07}~\cite{Charpiat08}~\cite{Liu08} have been proposed to reduce the user efforts required for image colorization, significantly. However, it still requires a considerable amount of manual efforts to generate satisfactory results for a broad range of images.

The colorization process involves assigning an appropriate three-dimensional value (RGB) to a pixel in a grayscale image by using only one-dimensional information(luminance or Intensity). The problem is ill-posed since several color values may have the same intensity value. Due to this reason, there is no unique solution for the colorization problem and the human interaction plays an important role in colorization process. 

Colorization methods can be roughly divided into two categories: interactive colorization methods and automatic colorization methods. The interactive colorization techniques~\cite{Levin04}~\cite{Irony05}~\cite{Huang05}~\cite{Yatziv06}~\cite{Luan07}~\cite{Alex11} require user input such as manually marked color scribbles or the pre-segmented regions. The color scribbles or the pre-segmented masks are used to provide the color information at high confidence image points and then these color values are used to spread the color into the whole grayscale image by using an optimization based framework. Instead of using user-specified color information, automatic colorization methods~\cite{Aaron01}~\cite{Welsh02}~\cite{Charpiat08}~\cite{Liu08} use one or more similar reference color images to transfer the color information to input grayscale image automatically. These algorithms either use the local image information of a pixel or a global optimization framework to automatically assign the suitable color value to each grayscale image pixel. 

In this paper, we proposed a learning-based color transfer algorithm which transfers the colors from one or more reference color images to the grayscale image without any user intervention. The algorithm uses the reference color images to learn the relationship between different image features and their corresponding color values. This information is then used to predict the possible color values of grayscale image pixels.  Unlike the previous techniques, which require dozens of carefully placed color scribbles~\cite{Levin04}~\cite{Yatziv06}~\cite{Luan07}, the partially segmented example color image~\cite{Irony05} or the segmented foreground and background grayscale image~\cite{Alex11}, the proposed colorization technique requires considerable less amount of user inputs which involve supplying one or more references color images only. 

Rather than working on an individual image pixel, the proposed algorithm uses the superpixel~\cite{Alex09} representation of input grayscale image and the reference color images which reduces the complexity of the algorithm by grouping the image pixels that exhibit similar image properties. Based on the local appearance of these superpixels and their neighboring superpixels, we compute a set of image features for each of these superpixels. We quantize the average color values of the reference color image superpixels to compute a color label for each of these superpixels. The image features computed for reference image superpixels and their corresponding color labels are then used to train a randomize decision forest. After the training, this randomized decision forest is used to predict the color labels of input gray image superpixels. While transferring the color values corresponding to the color labels, we transfer only chromaticity values as a \emph{micro-scribble} at the centroid of the grayscale image superpixels and then refine these \emph{micro-scribbles} by using a voting-based approach. To smoothly spread these \emph{micro-scribbles} across the superpixel, we use an optimization-based colorization technique to generate the final colorized result.

The rest of paper is structured as follows: In Section 2, we briefly cover the related work; Section 3 describes the proposed colorization algorithm. Section 4 presents the colorization results and a detailed comparison with other state-of-the-art colorization algorithms. Section 5 concludes our paper.

\begin{figure*}\center
\begin{minipage}[t]{\linewidth}
\centering
\begin{tabular}{c@{\hspace{0mm}}}
\\\vspace{0mm}\\
\includegraphics[width=1\linewidth]{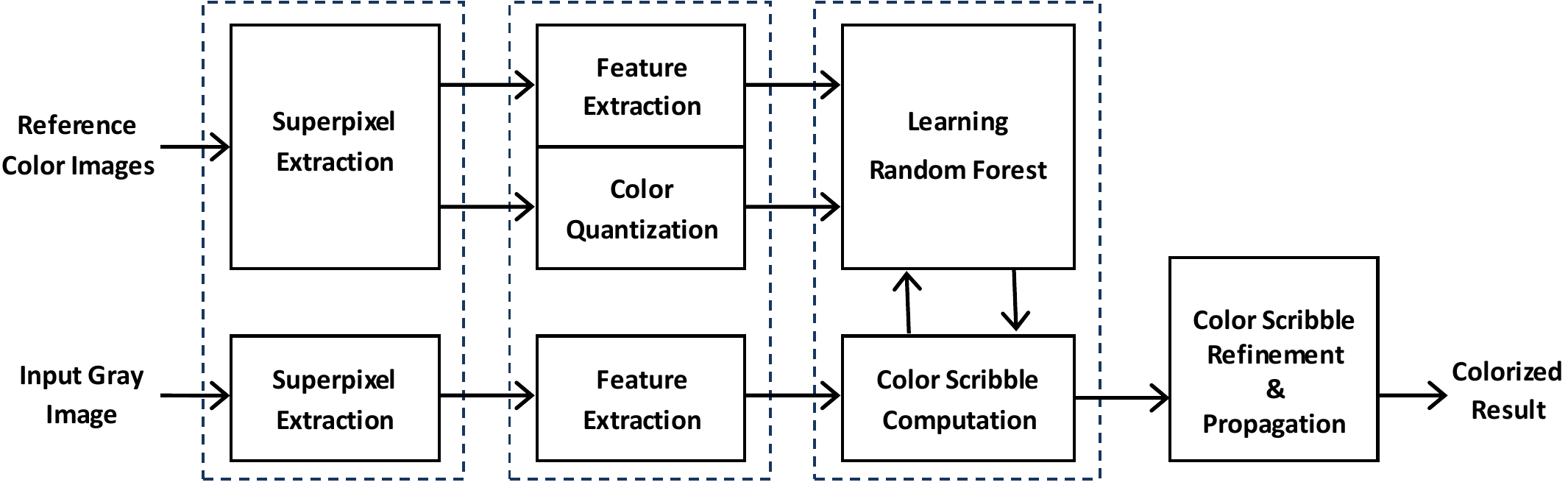}
\end{tabular}
\end{minipage}
 \caption{Block diagram of the proposed algorithm.}
\label{fig:pipeline}
\end{figure*}

\section{Related Work}
\label{sec:2}
To transfer the color to grayscale images, a number of colorization techniques have been proposed recently. Levin $et\ al.$~\cite{Levin04} proposed a simple yet very effective colorization algorithm that requires the user to provide color scribbles at various image regions. These color scribbles are propagated automatically by using a least-squares optimization method to produce the final colorization results. Huang $et\ al.$~\cite{Huang05} improved on this method by using the edge information and reduce the color blending at object boundaries. Yatziv and Sapiro~\cite{Yatziv06} used multiple scribbles to compute the color value of a pixel. Luan $et\ al.$~\cite{Luan07} reduced the required number of color scribbles significantly by using the texture information. Although these algorithms are quite efficient, it require considerable amount of user efforts and the colorization results totally depend on the artistic skills of the user. 

To automate the colorization process, Welsh $et\ al.$~\cite{Welsh02} proposed an algorithm that colorizes a grayscale image by matching small neighborhood of each pixel to the reference color image pixels. The algorithm transfers the chromaticity information of the best match to input gray image pixels. This approach is also one of the applications of image analogies framework~\cite{Aaron01} which learns the relationships between the grayscale image and the reference color image pixels. This framework deals with only a few colors and the colorization results contain a lot of small artifacts due to the lack of any spatial coherency criterion during the colorization process.

Irony $et\ al.$~\cite{Irony05} addressed the spatial coherency problem by using the partially segmented input reference color image. The algorithm learns the color and texture information from the partially segmented image region and then uses this learned information to automatically segment the grayscale image into locally homogeneous regions. The colors are then transferred to few image points in these segmented image regions where the prediction reaches to the highest confidence. These image points are then considered as the color scribbles and the colors are spread across the entire image by using an optimization-based approach~\cite{Levin04}. This algorithm improves the colorization results due to its use of spatial coherency criterion; however it requires additional inputs from the users.
 
Charpiat $et\ al.$~\cite{Charpiat08} formulated the colorization problem as an optimization problem and solved it by using a graph cuts based approach that automatically assigns colors to grayscale image pixels by using an energy minimization function. The algorithm does not require any user intervention during the colorization process, however the colorization results heavily depend on a number of input parameters. The algorithm requires an image based parameter tuning to generate good colorization results.

Liu $et\ al.$~\cite{Liu08} proposed an illumination insensitive colorization technique that uses the Internet images to colorize the grayscale image. The limitation of their method is that it requires the color reference images that have been taken from similar viewing angle as of the input gray image which restricts the applicability of this method to the images that contain non-deformable objects (mainly landmarks). 

Recently, Chia $et\ al.$~\cite{Alex11} proposed a belief propagation based approach to colorize the grayscale image. Their algorithm also leverages on the rich image content available on the Internet. The algorithm works quite well on a broad category of images; however, It requires substantial amount of user intervention during the colorization process where the user needs to provide the segmented masks for all major foreground objects and a semantic text label to describe each of these foreground objects.

\section{Algorithm}
\label{sec:3}
The proposed algorithm uses a learning-based approach to colorize gray input images by using one or more user-supplied reference color images. An overview of our colorization method is shown in Figure \ref{fig:pipeline}. As shown in the figure, the proposed colorization method comprises four key steps: (a) superpixel extraction (b) feature extraction, (c) feature learning and (d) color scribble refinement and propagation. Each of these steps is described below. 
\subsection{Superpixel extraction}
To reduce the complexity of the algorithm and enforce higher spatial consistency, rather than working on an individual image pixel, the proposed algorithm works at the resolution of superpixel. The superpixels are the compact representation of an image which consists of a group of connected pixels that exhibit similar image properties. To compute the superpixels, we used a geometric-flow based approach proposed by Levinshtein $et\ al.$~\cite{Alex09}. The algorithm produces the superpixels of uniform size and shape and also preserves the original image edges. In our experiments, we used the input parameters \emph{time step} and \emph{maximum number of iterations} as 0.5 and 500, respectively. These values are the default values provided by the authors along with their source code\footnote{http://www.cs.toronto.edu/~babalex/research.html}. Depending on the image size, the \emph{number of input superpixels} is chosen to keep the average size of output superpixels around 40 pixels.

\subsection{Feature extraction}
After extracting the superpixels from input gray image and reference color images, we compute a set of feature vectors for each of these superpixels. We compute these image features based on their local as well as their neighboring superpixels as follows:

\subsubsection{Intensity feature}
We compute a two-dimensional intensity feature vector for each superpixel by using their intensity value. The first dimension of the feature vector contains an average intensity value of the superpixel computed by using the intensity values of all the pixels within that superpixel. The second dimension of the feature vector contains an average intensity value of its connected neighboring superpixels.

\subsubsection{Standard deviation feature}
Similar to intensity feature vector, we compute a two-dimensional feature vector based on the standard deviation values computed in small neighborhood around each image pixel. For all experiments in this paper, we used a $5$ x $5$ neighboring size to compute the standard deviation at each image pixel. We use this standard deviation value computed at each pixel to calculate the feature vector in the same way as that computed for the intensity feature vectors. 

\subsubsection{Gabor feature}
 In addition to the standard deviation feature vector, we use the Gabor filters~\cite{BS96} to exploit the image texture features. We compute a 40-dimensional feature vector at each image pixel by applying Gabor filters with eight orientations varying in increments of $\pi/8$ from 0 to $7\pi/8$, at five exponential scales $exp(i \times \pi), I = 0, 1, 2, 3, 4$. The feature vectors of all pixels within a superpixel are grouped together and their mean value along each dimension is used to form a  40-dimensional feature vector of a superpixel.  

\subsubsection{Dense SIFT feature}
We compute a 128-dimensional dense SIFT descriptors~\cite{Liu11} at each image pixel. To compute the dense SIFT descriptor of a pixel, we divide its neighborhood pixels into a $4$ x $4$ cell array. For each of these cells, we quantize the orientation into 8 bins to obtain $4$ x $4$ x $8$ $=\ 128$ dimensional feature descriptor. After computing the SIFT descriptor for each image, the descriptor of all the pixels within a superpixel are grouped together to form a 128-dimensional feature vector of a superpixel.

\subsection{Feature learning}
This section describes the two components of learning process: color quantization of the reference color image superpixels to compute the labels for reference color image superpixels, and the randomized decision forest trained by using the superpixel feature vectors and their corresponding color labels. Each of these components is discussed below.

\subsubsection{Color quantization}
\label{sec:quantization}
To learn the random forest, we will need to quantize the color space to generate a label for each of the reference color image superpixels. We use the CIEL$ab$ color space to discretize the color values due to its proximity to human vision system. The average chromatic values $a$ and $b$ of all the pixels within a superpixel are used to compute a label for that superpixel. We compute these average chromatic values of all reference color image superpixels and cluster them by using the $k$-means algorithm. Each cluster is represented by the mean $a$ and $b$ values of the superpixels in it and the cluster index is used to represent the color label of the superpixel.

We use the feature vectors of reference color image superpixels and their corresponding color labels to learn the randomized decision forest. After learning, we use this decision forest to predict the color labels of input gray image superpixels. 

\subsubsection{Randomized decision forest}
Randomized decision trees and forests~\cite{Leo01} are used effectively as multi-class classifier in many computer vision applications~\cite{Anna07}~\cite{Gang11}~\cite{Shotton11}. They are proven fast classifier~\cite{Toby08} and can be implemented on GPU efficiently. Randomized decision forest consists of $T$ decision trees, where each tree grown by using some form of randomization. Each internal node of the tree contains a feature $f_{n}$ and a threshold $\tau$ that best splits the data space to be classified. To classify a superpixel $s$ in a image I, we start from the root node and traverse the tree to either left or right according to the threshold value $\tau$ stored at that node. At the leaf node of a tree $t$, a learned distribution $P_{t}(c|I, s)$ over the color labels $c$ is stored. The distributions of all the trees in a forest, are averaged together to the final classification of a superpixel:
\begin{equation}
\hspace{15mm} P(c|I, s) = \frac{1}{T}\sum_{t=1}^{T}{P_{t}(c|I, s)}.
\end{equation}
To train each decision tree $t$, we use the following algorithm~\cite{Shotton11}:\\\\
1. Randomly select a set of splitting candidates $\chi$ $=$ $(n, \tau)$, where $n$ and $\tau$ represent the dimensions of feature vector and thresholds respectively.\\\vspace{-2.5mm}\\
2. Divide the set of examples $E = {(I, s)}$ into two subsets $L$ and $R$ by using each $\chi$:
\begin{equation}
\hspace{15mm} E_{l}(\chi) = {(I, s)|f_{n}(I, s) < \tau}
\end{equation}
\begin{equation}
\hspace{15mm} E_{r}(\chi) = {(I, s)|f_{n}(I, s) \geq \tau}
\end{equation}
3. Compute the $\chi$ that offers the maximal information gain at the node:
\begin{equation}
\hspace{20mm} \bar{\chi} = \operatorname*{arg\,max}_\chi G(\chi)
\end{equation}
\begin{equation}
\hspace{8mm} G(\chi) = H(E) - \sum_{q \in {l, r}}{\frac{|E_{q}(\chi)|}{|E|}H(E_{q}(\chi))}
\end{equation}
where $H(E)$ is the Shannon entropy which is computed on all $(I, s) \in E$.\\\vspace{-2.5mm}\\
4. If the largest gain $G(\bar{\chi})$ is sufficient, then recurse for left and right subsets $E_{l}(\bar{\chi})$ and $E_{r}(\bar{\chi})$ till node encounters no more examples.\\\vspace{-2.5mm}\\
In our experiments, the randomized decision forest consists of $1500$ decision trees, $172$ $(2+2+40+128)$ candidate features $(n)$ and $13$ candidate thresholds $(\tau)$.

\begin{figure*}\center
\begin{minipage}[t]{\linewidth}
\centering
\begin{tabular}{c@{\hspace{.7mm}} c@{\hspace{0.7mm}} c@{\hspace{0.7mm}} c@{\hspace{0.7mm}} c@{\hspace{0.7mm}} |c@{\hspace{0mm}}}
\includegraphics[width=0.158\linewidth]{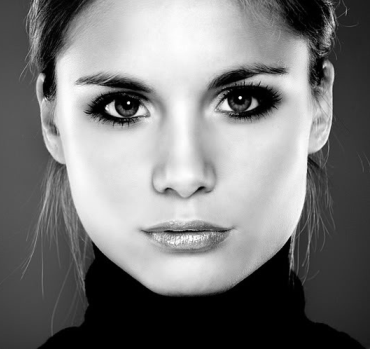}&
\includegraphics[width=0.158\linewidth]{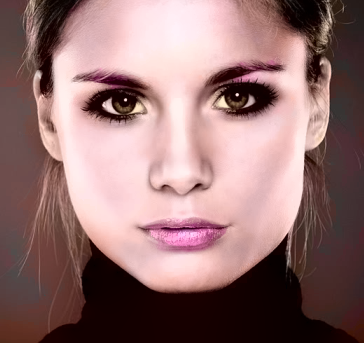}&
\includegraphics[width=0.158\linewidth]{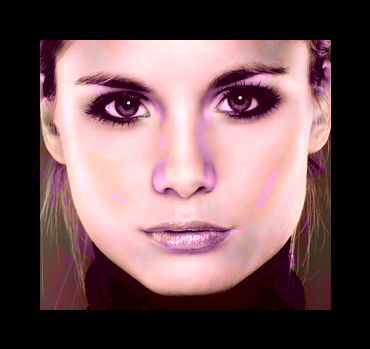}&
\includegraphics[width=0.158\linewidth]{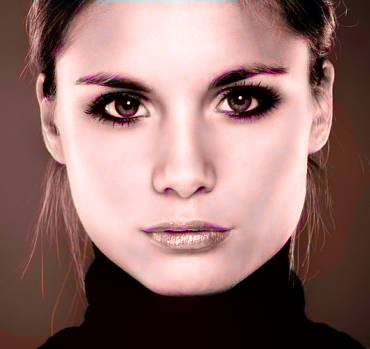}&
\includegraphics[width=0.158\linewidth]{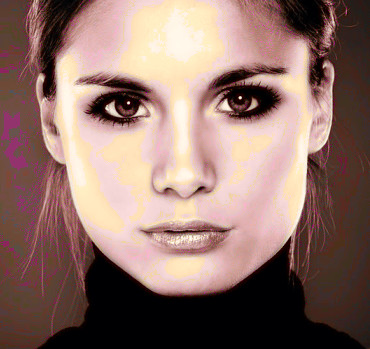}&
\includegraphics[width=0.1\linewidth]{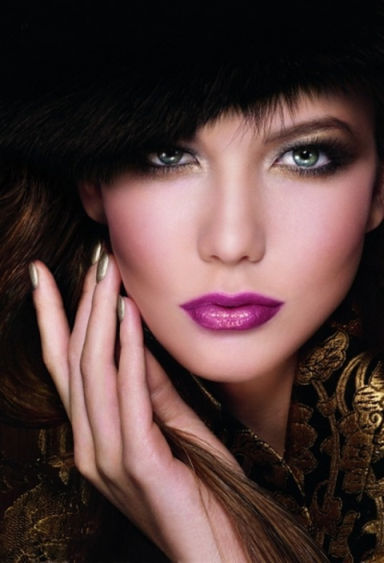}\\
\includegraphics[width=0.158\linewidth]{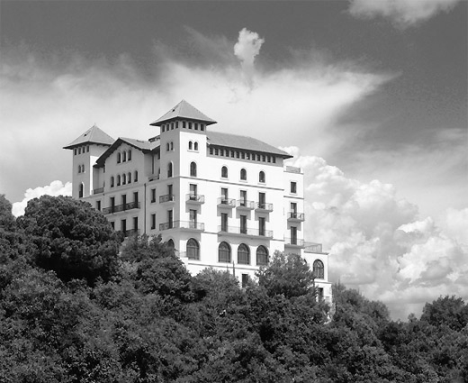}&
\includegraphics[width=0.158\linewidth]{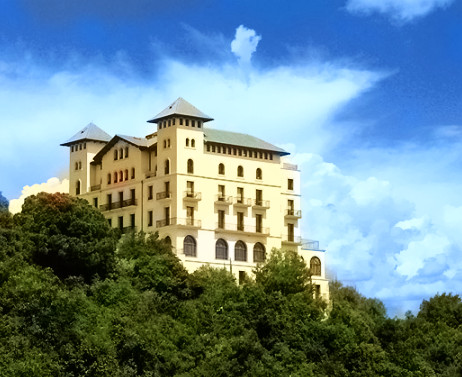}&
\includegraphics[width=0.158\linewidth]{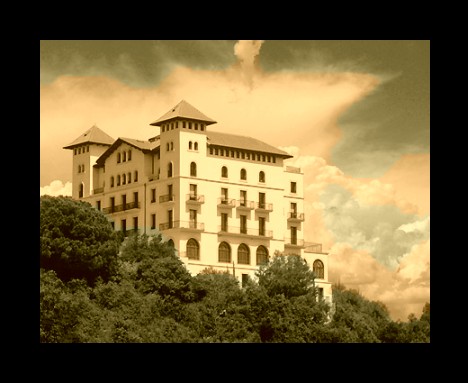}&
\includegraphics[width=0.158\linewidth]{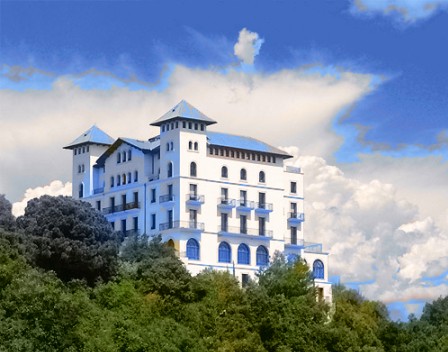}&
\includegraphics[width=0.158\linewidth]{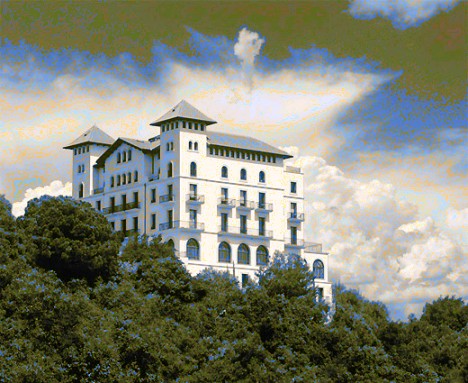}&
\includegraphics[width=0.158\linewidth]{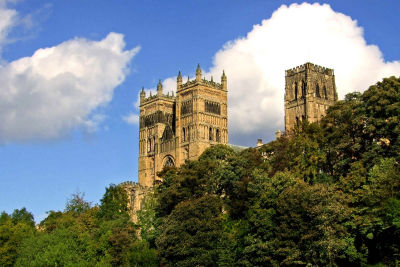}\\
\includegraphics[width=0.158\linewidth]{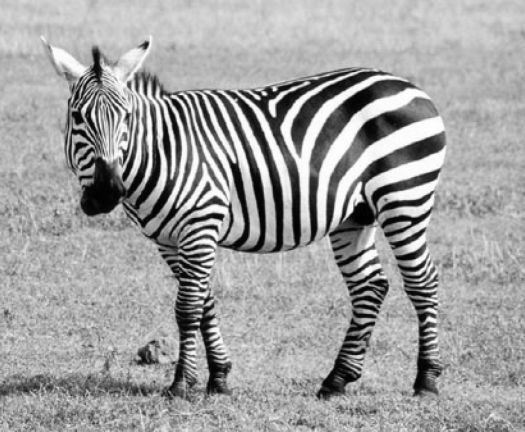}&
\includegraphics[width=0.158\linewidth]{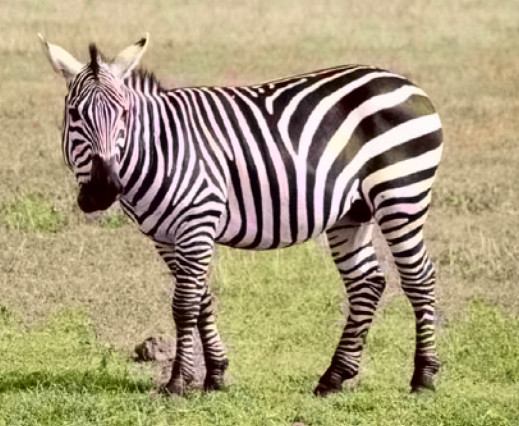}&
\includegraphics[width=0.158\linewidth]{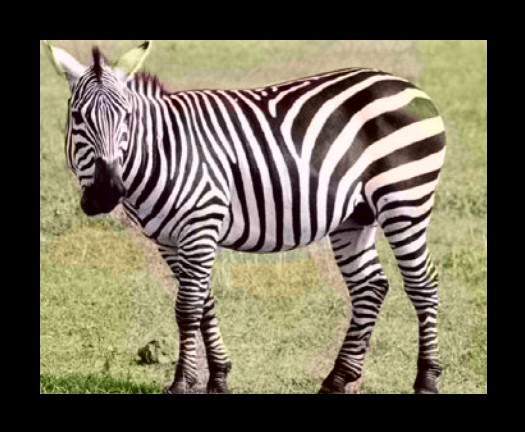}&
\includegraphics[width=0.158\linewidth]{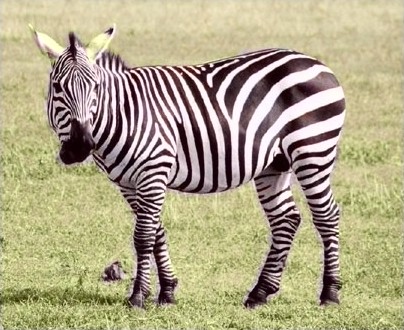}&
\includegraphics[width=0.158\linewidth]{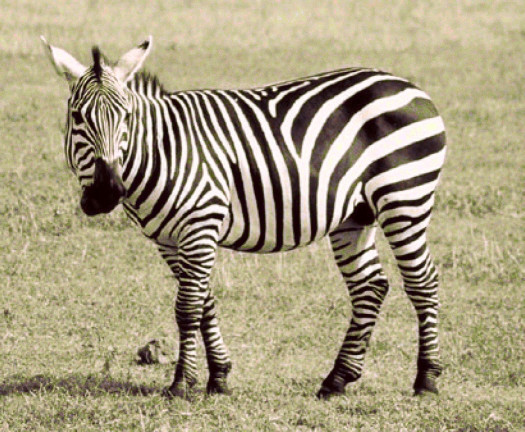}&
\includegraphics[width=0.158\linewidth]{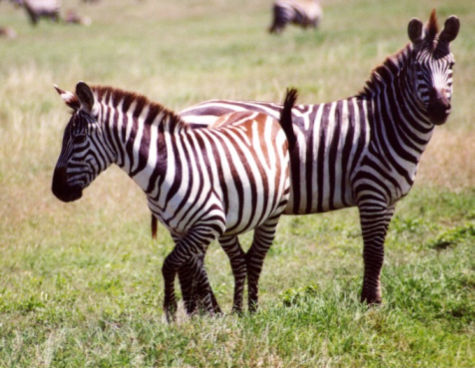}\\
\includegraphics[width=0.158\linewidth]{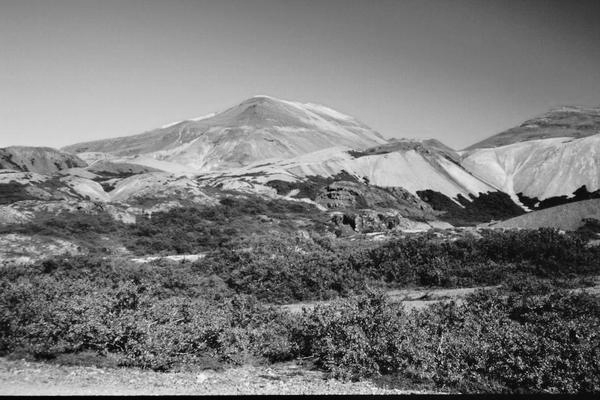}&
\includegraphics[width=0.158\linewidth]{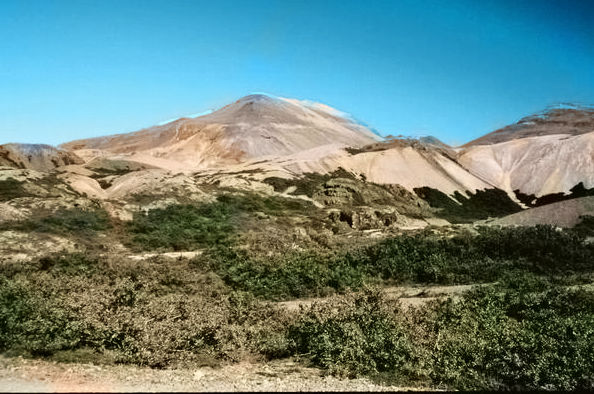}&
\includegraphics[width=0.158\linewidth]{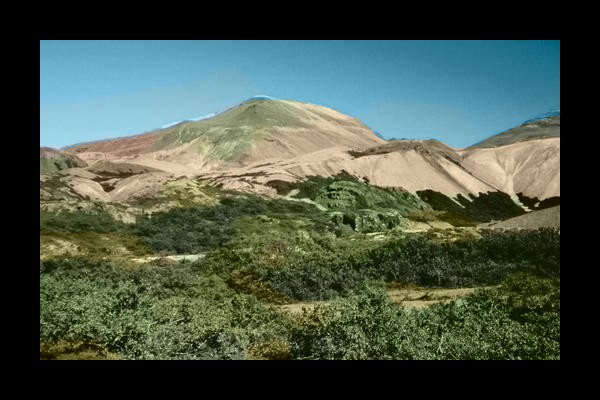}&
\includegraphics[width=0.158\linewidth]{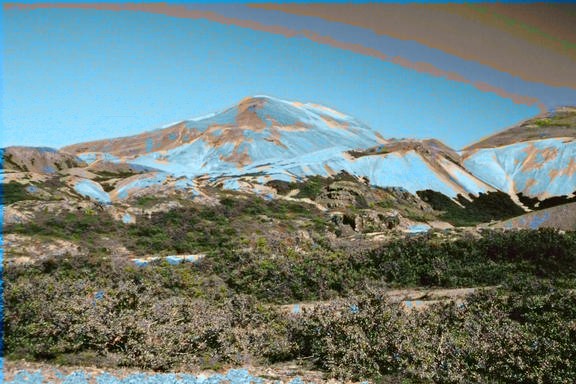}&
\includegraphics[width=0.158\linewidth]{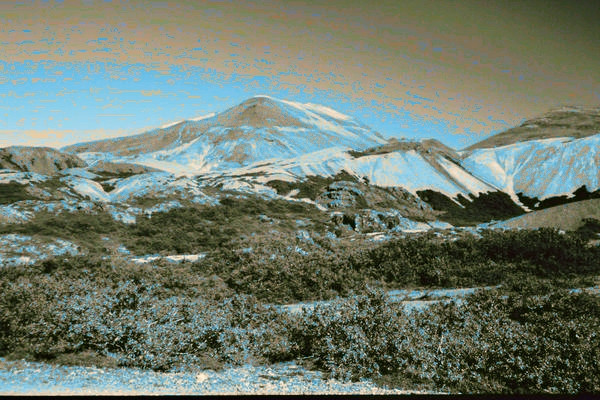}&
\includegraphics[width=0.158\linewidth]{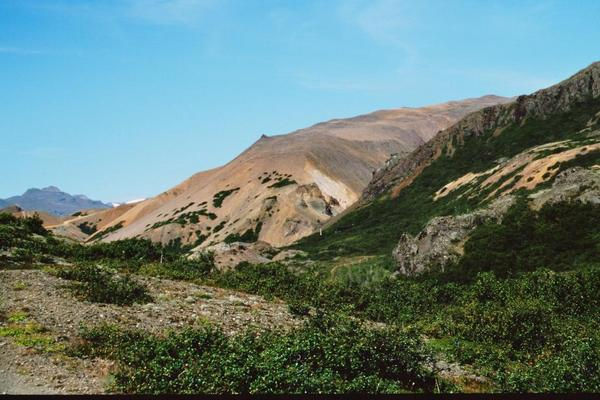}\\
\includegraphics[width=0.158\linewidth]{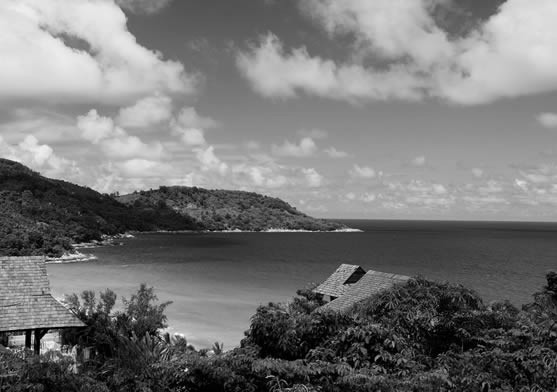}&
\includegraphics[width=0.158\linewidth]{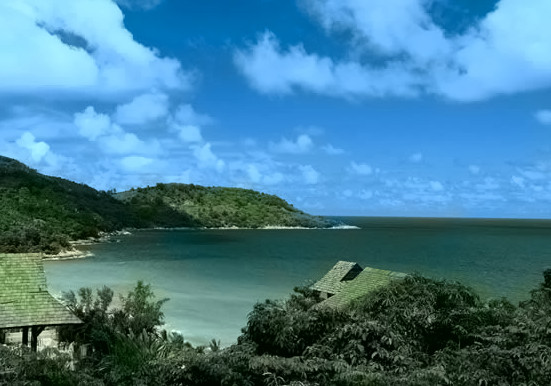}&
\includegraphics[width=0.158\linewidth]{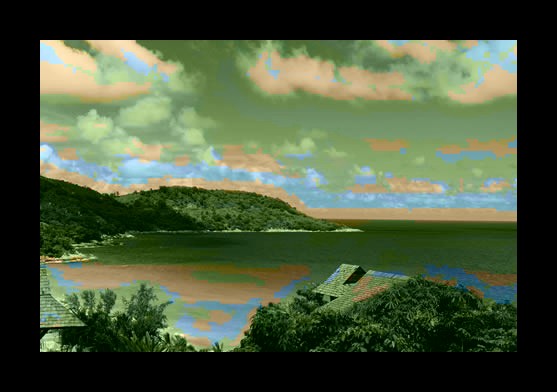}&
\includegraphics[width=0.158\linewidth]{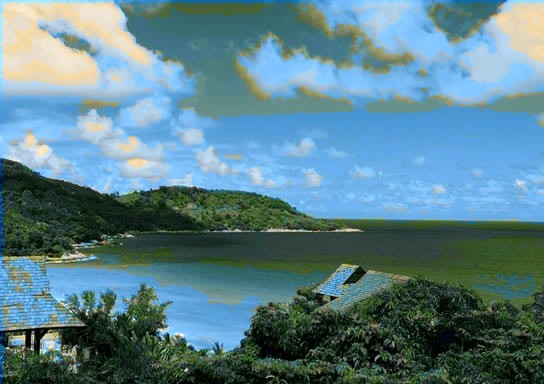}&
\includegraphics[width=0.158\linewidth]{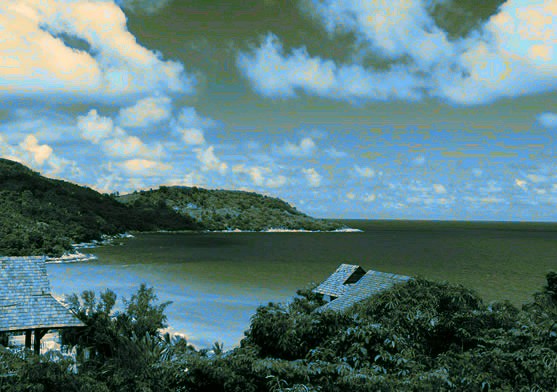}&
\includegraphics[width=0.147\linewidth]{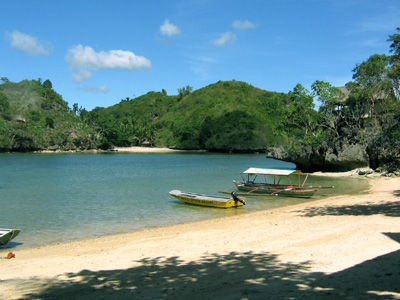}\\
\includegraphics[width=0.158\linewidth]{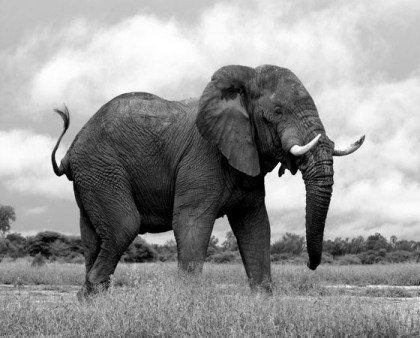}&
\includegraphics[width=0.158\linewidth]{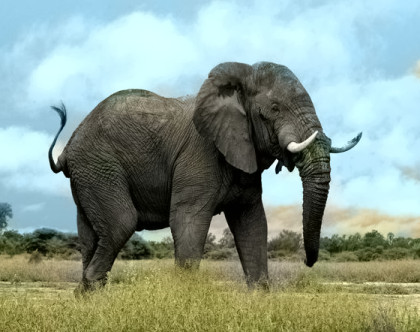}&
\includegraphics[width=0.158\linewidth]{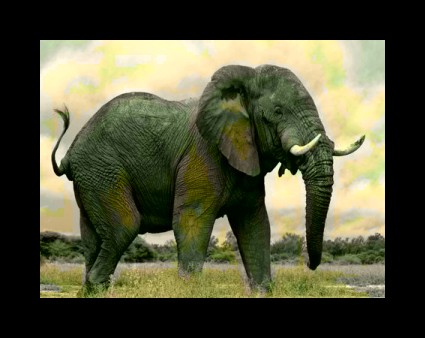}&
\includegraphics[width=0.158\linewidth]{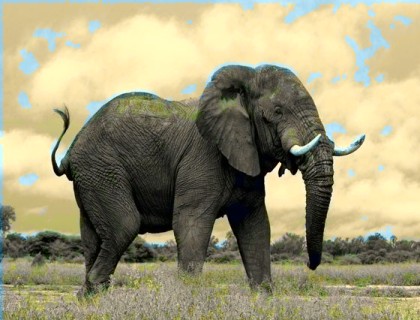}&
\includegraphics[width=0.158\linewidth]{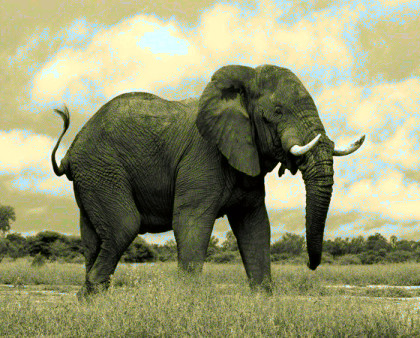}&
\includegraphics[width=0.158\linewidth]{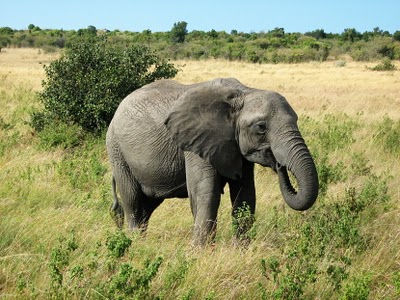}\\
\includegraphics[width=0.158\linewidth]{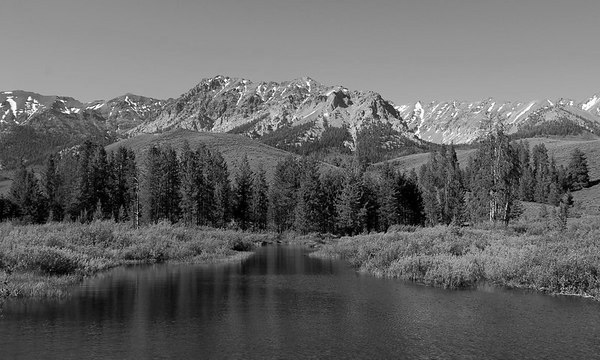}&
\includegraphics[width=0.158\linewidth]{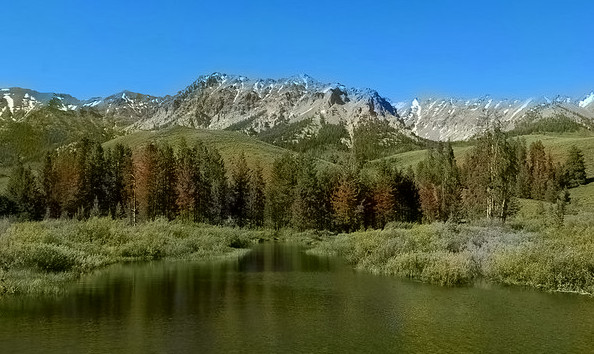}&
\includegraphics[width=0.158\linewidth]{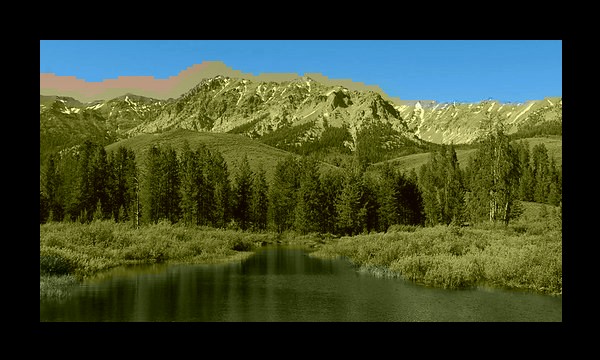}&
\includegraphics[width=0.158\linewidth]{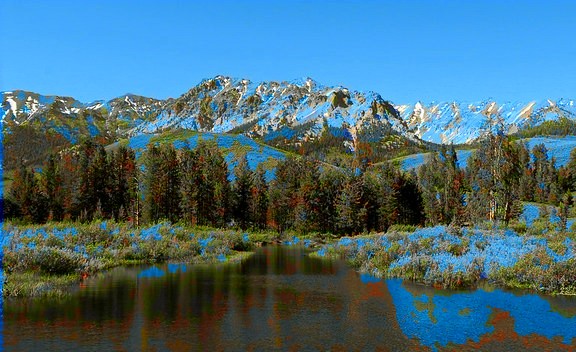}&
\includegraphics[width=0.158\linewidth]{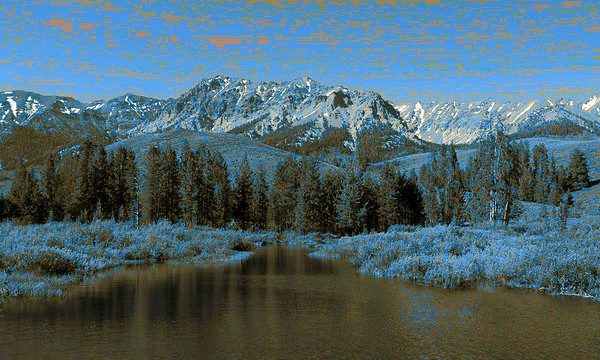}&
\includegraphics[width=0.122\linewidth]{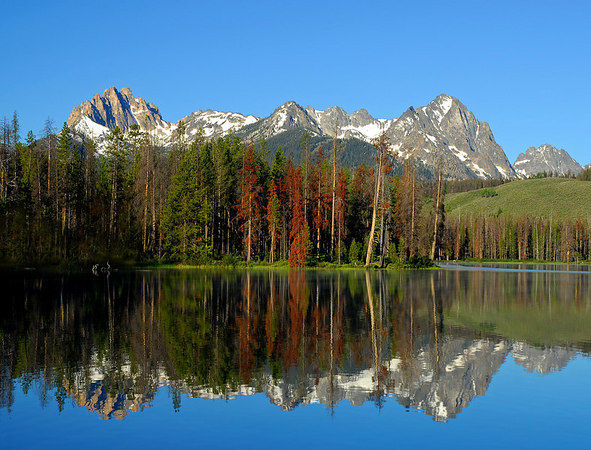}\\
Input gray & Our result & Charpiat $et\ al.$~\cite{Charpiat08} & Irony $et\ al.$~\cite{Irony05} & Welsh $et\ al.$~\cite{Welsh02} & Reference image
\end{tabular}
\end{minipage}
 \caption{Comparison with existing state-of-the-art colorization
 methods. The last column shows the reference color images that
have been used by all algorithms for colorizing the input grayscale images.}
\label{fig:ECCV_SIGGRAPH}
\end{figure*}

\subsection{Color scribble refinement and propagation}
\label{sec:ColorRefinement}
The color labels of input gray image superpixels by randomized decision forest are often less reliable near the object boundaries. To improve the accuracy in such image regions, we use a voting-based approach that reduces the number of such superpixels. To perform the voting, we perform automatic image segmentation~\cite{Meer02} on input gray image. We used the source code\footnote{http://coewww.rutgers.edu/riul/research/code/EDISON/} provided by authors with the input parameters \emph{SpatialBandwidth} and \emph{RangeBandwidth} as $2$ and $3$, respectively. For each image segment, we check for the color labels that have been assigned to the superpixels within that image segment. We assume that the connected superpixels that are the part of an image segment, will also have the same color label. The color label which belongs to majority of superpixels is then used to update the color labels of other superpixels in that image segment. 

After updating the color labels of all the superpixels, we use these color labels to generate the color scribble for each superpixel. As we mentioned earlier (in Section \ref{sec:quantization}), each color label represents the chromatic color values $a$ and $b$. We use these chromatic color values to generate a \emph{micro-scribble} at the centroid of the gray image superpixel. While assigning the chromatic color values to the superpixels, we do not modify the luminance channel ($L$) of the input gray image. These \emph{micro-scribbles} are then used to propagate the color values to complete input gray image. We use the approach proposed by Levin $et\ al.$~\cite{Levin04} to propagate the color values across all image pixels. The algorithm is based on the principle that the neighboring pixels that have similar luminance value should also have similar colors. The algorithm works in $YUV$ color space and attempts to minimize the difference $J(C)$ between the color assigned to a pixel $p$ and the weighted average of the colors assigned to its neighboring pixels.
\begin{equation}
\hspace{8mm}J(C) = \sum_{p\in I}\left(C(p)- \sum_{q\in N(p)} w_{pq}\ C(q) \right)^2,\end{equation}
where $q$ represents the neighboring pixels of $p$ and the weights $w_{pq}$ are determined by the similarity of their luminance ($Y$).
\begin{equation}
\hspace{20mm}w_{pq} \propto e^{-(Y(p)-Y(q))^2/2\sigma_{p}^2} .
\end{equation}
For more details of this algorithm, the readers can refer to~\cite{Levin04}.

\begin{figure*}\center
\begin{minipage}[t]{\linewidth}
\centering
\vspace{-0.2cm}
\begin{tabular}{c@{\hspace{0.5mm}} c@{\hspace{0.5mm}} c@{\hspace{0.5mm}} c@{\hspace{0.5mm}} c@{\hspace{0.5mm}} c@{\hspace{0.5mm}} c@{\hspace{0mm}}}
\multicolumn{7}{c}{\includegraphics[height=0.126\linewidth]{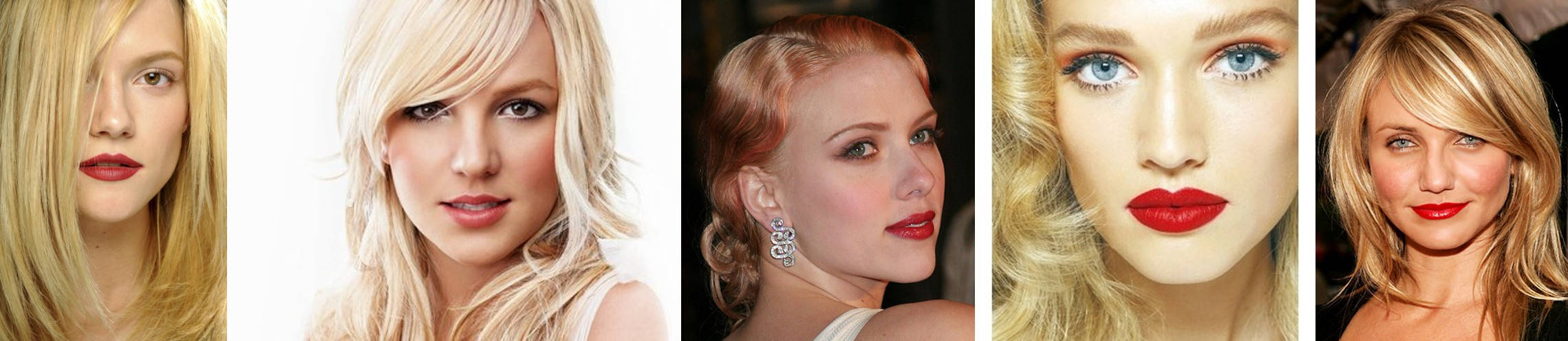}}
\\
\includegraphics[height=0.126\columnwidth]{images/port30.png}&
\includegraphics[height=0.126\columnwidth]{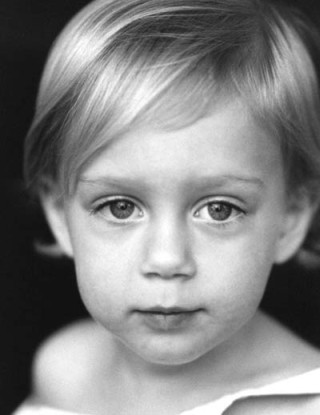}&
\includegraphics[height=0.126\columnwidth]{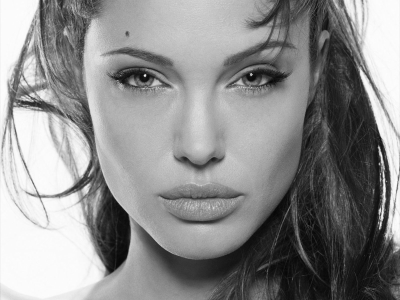}&
\includegraphics[height=0.126\columnwidth]{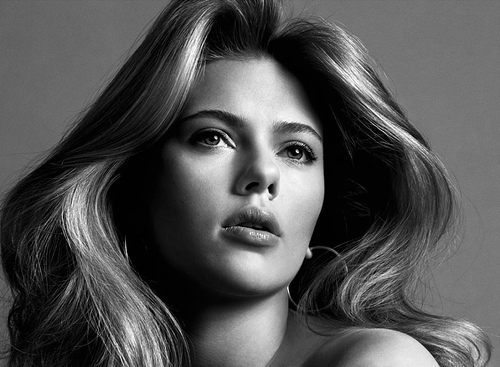}&
\includegraphics[height=0.126\columnwidth]{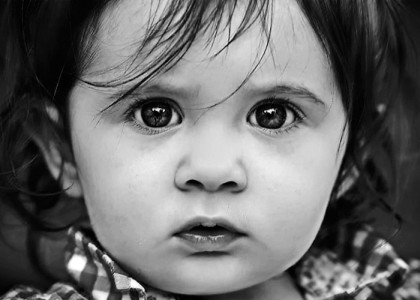}&
\includegraphics[height=0.126\columnwidth]{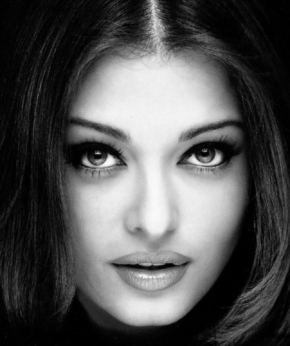}&
\includegraphics[height=0.126\columnwidth]{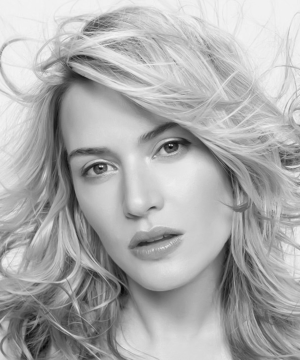}
\\
\includegraphics[height=0.126\columnwidth]{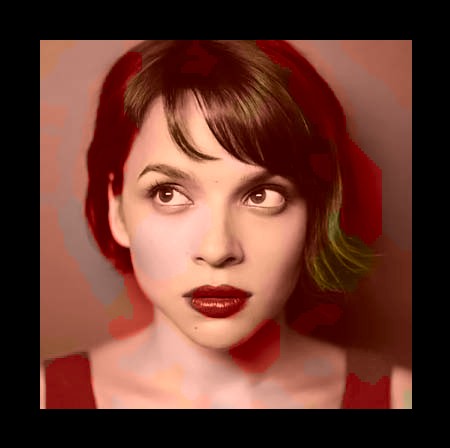}&
\includegraphics[height=0.126\columnwidth]{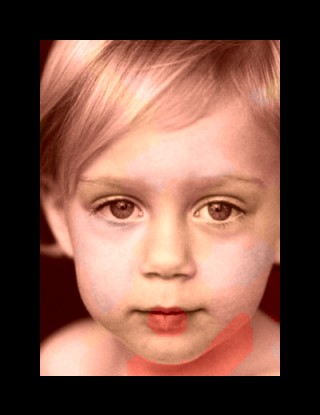}&
\includegraphics[height=0.126\columnwidth]{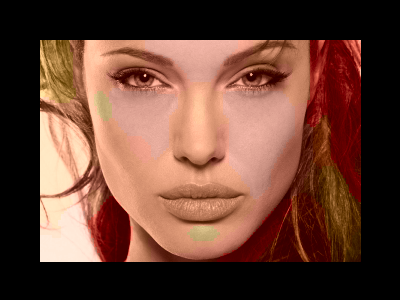}&
\includegraphics[height=0.126\columnwidth]{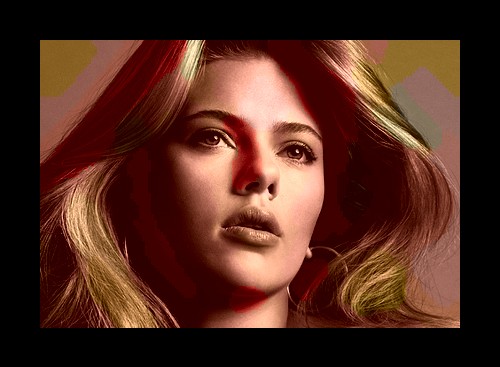}&
\includegraphics[height=0.126\columnwidth]{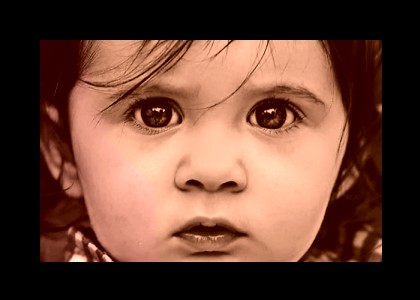}&
\includegraphics[height=0.126\columnwidth]{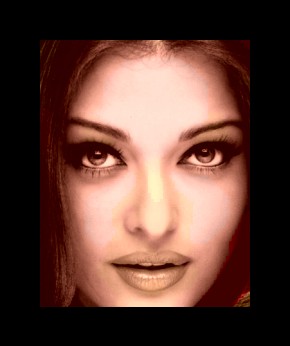}&
\includegraphics[height=0.126\columnwidth]{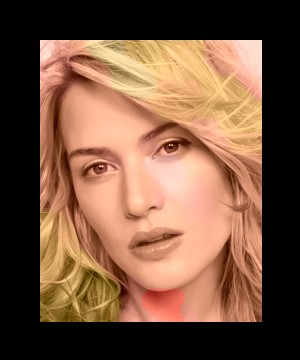}
\\
\includegraphics[height=0.126\columnwidth]{images/User/port30_Result_Sat.jpg}&
\includegraphics[height=0.126\columnwidth]{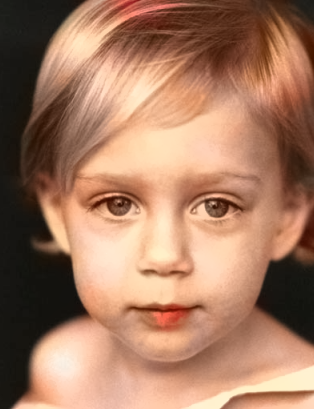}&
\includegraphics[height=0.126\columnwidth]{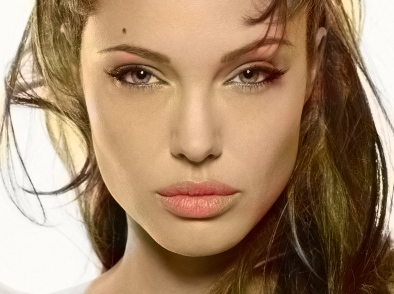}&
\includegraphics[height=0.126\columnwidth]{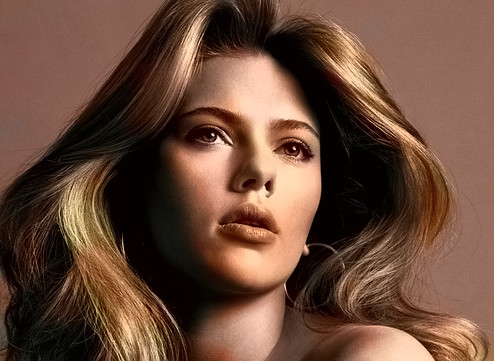}&
\includegraphics[height=0.126\columnwidth]{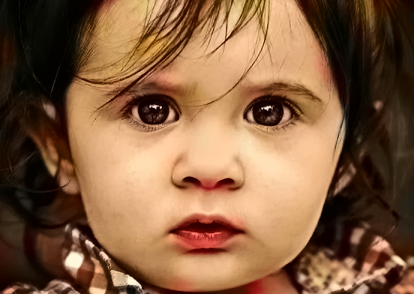}&
\includegraphics[height=0.126\columnwidth]{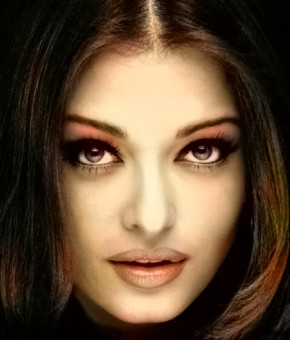}&
\includegraphics[height=0.126\columnwidth]{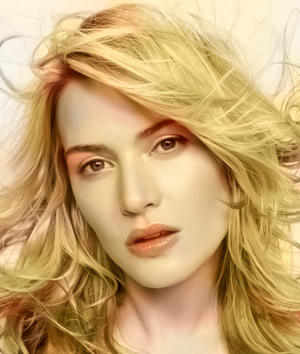}
\end{tabular}
\end{minipage}
 \caption{Comparison with another learning-based colorization algorithm~\cite{Charpiat08} on portrait images. The first row shows the reference color images used by both colorization algorithms. The input gray images have been shown in second row. The third and fourth rows show the colorization results obtained by using Charpiat $et\ al.$'s method~\cite{Charpiat08} and the proposed algorithm, respectively.}
\label{fig:Portraits}
\end{figure*}

\begin{figure*}\center
\begin{minipage}[t]{\linewidth}
\centering
\vspace{0.7mm}
\begin{tabular}{c@{\hspace{0.8mm}} |c@{\hspace{0.8mm}} c@{\hspace{0.8mm}} c@{\hspace{0.8mm}} |c@{\hspace{0mm}}}
\includegraphics[width=0.187\linewidth]{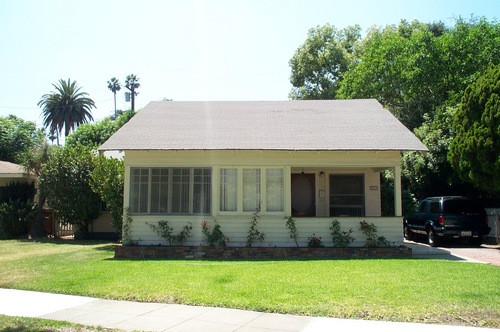}&
\includegraphics[width=0.187\linewidth]{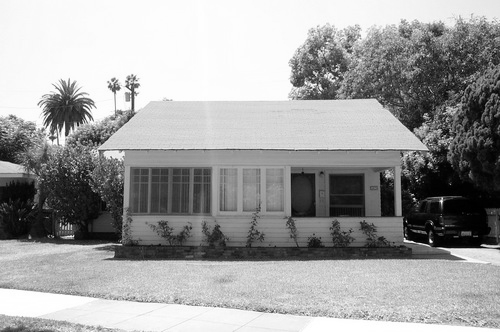}&
\includegraphics[width=0.187\linewidth]{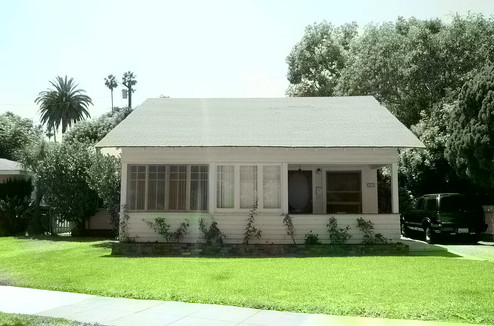}&
\includegraphics[width=0.187\linewidth]{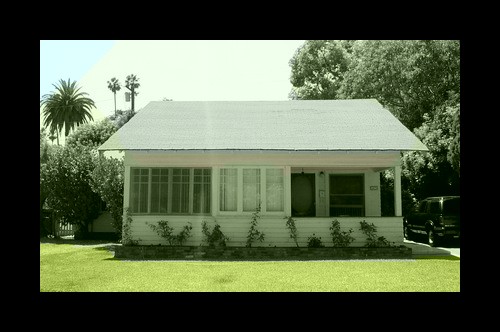}&
\includegraphics[width=0.187\linewidth]{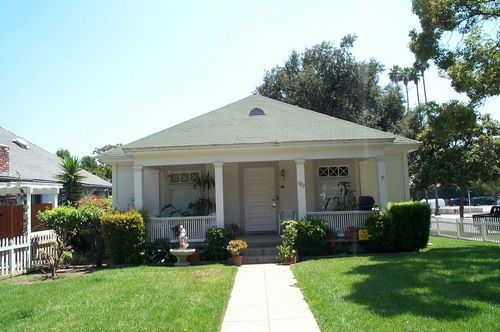}\\
\includegraphics[width=0.187\linewidth]{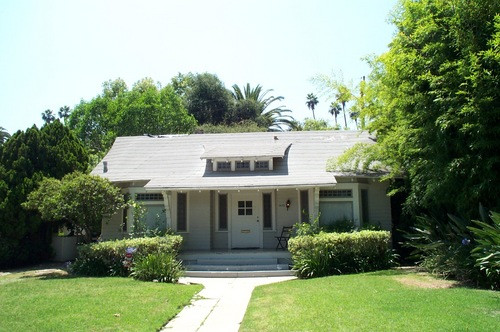}&
\includegraphics[width=0.187\linewidth]{images/maison1.png}&
\includegraphics[width=0.187\linewidth]{images/maison1_Result.jpg}&
\includegraphics[width=0.187\linewidth]{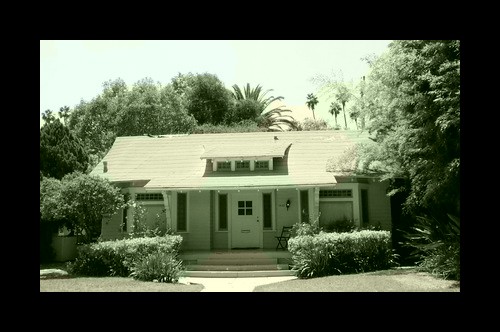}&
\includegraphics[width=0.187\linewidth]{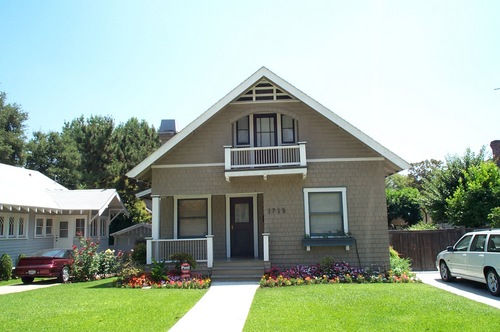}\\
Ground truth & Input image & Our results & Charpiat $et\ al.$~\cite{Charpiat08} &  Reference images
\end{tabular}
\end{minipage}
 \caption{Comparison with another learning-based colorization approach~\cite{Charpiat08}. The first and second columns show the ground truth color images and their corresponding grayscale images that have been used for colorization. The colorization results obtained by using our method and Charpiat $et\ al.$~\cite{Charpiat08} are shown in third and fourth columns, respectively. The last column contains the reference color images that have been used for training by both algorithms.}
\label{fig:Maison}
\end{figure*}

\begin{figure*}\center
\begin{minipage}[t]{\linewidth}
\centering
\vspace{0.7mm}
\begin{tabular}{c@{\hspace{0.8mm}} c@{\hspace{0.8mm}} c@{\hspace{0.8mm}} c@{\hspace{0.8mm}} c@{\hspace{0.8mm}} c@{\hspace{0.8mm}}}
\includegraphics[height=0.207\linewidth]{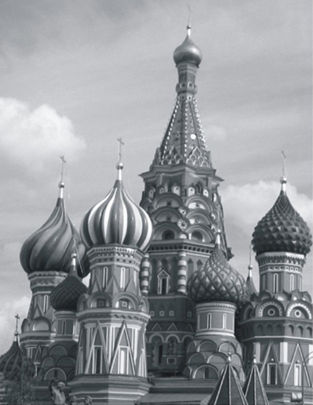}&
\includegraphics[height=0.207\linewidth]{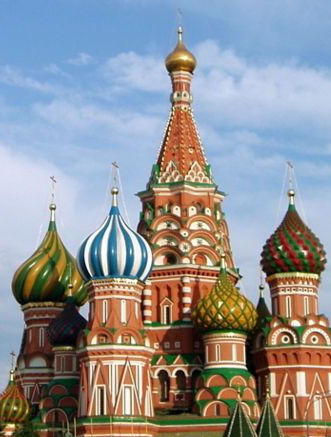}&
\includegraphics[height=0.207\linewidth]{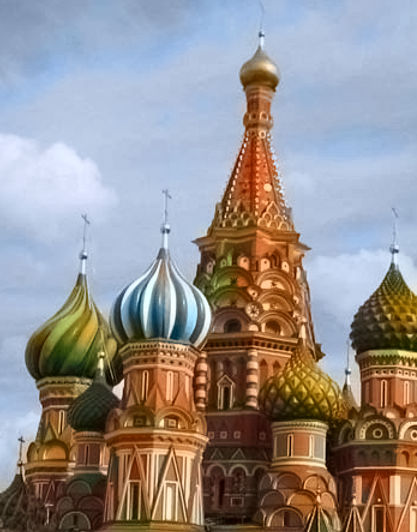}&
\includegraphics[height=0.207\linewidth]{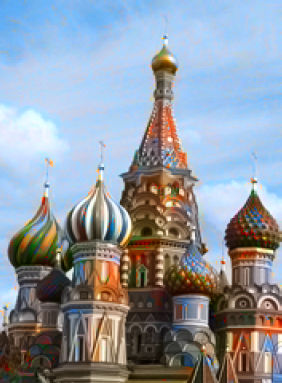}&
\includegraphics[height=0.207\linewidth]{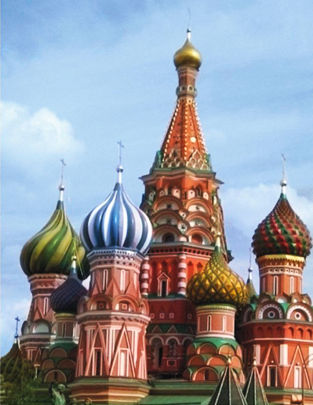}&
\includegraphics[height=0.207\linewidth]{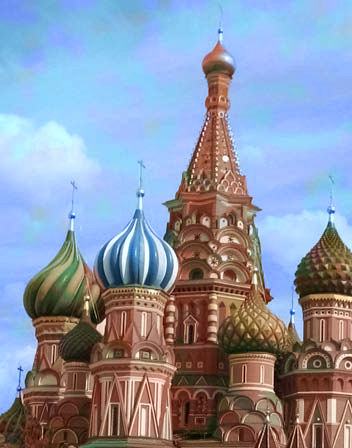}
\\
{\small (a)} & {\small (b)} & {\small (c)} & {\small (d)} & {\small (e)} & {\small (f)}
\end{tabular}
\end{minipage}
 \caption{\label{fig:SIG2008}
    Comparison with existing state-of-the-art colorization
 methods. (a) and (b) show the input gray image and reference 
color image used for color transfer. (c) shows our colorization 
result. (d-f) shows the colorization results obtained by using 
Irony et al.~\cite{Irony05}, Liu et al.~\cite{Liu08} and Chia 
et al.~\cite{Alex11}, respectively.}
\end{figure*}

\begin{figure*}\center
\begin{minipage}[t]{\linewidth}
\centering
\begin{tabular}{c@{\hspace{0.8mm}} c@{\hspace{0.8mm}} c@{\hspace{0.8mm}} c@{\hspace{0.8mm}} c@{\hspace{0.8mm}} c@{\hspace{0.8mm}}}
\includegraphics[height=0.16\linewidth]{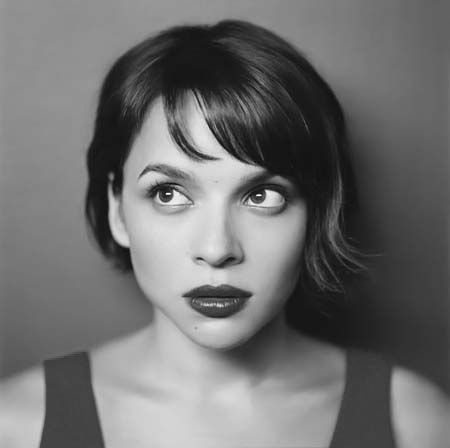}&
\includegraphics[height=0.16\linewidth]{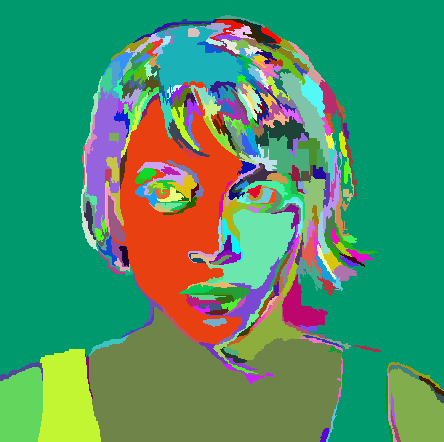}&
\includegraphics[height=0.16\linewidth]{images/User/port30_Result_Sat.jpg}&
\includegraphics[height=0.16\linewidth]{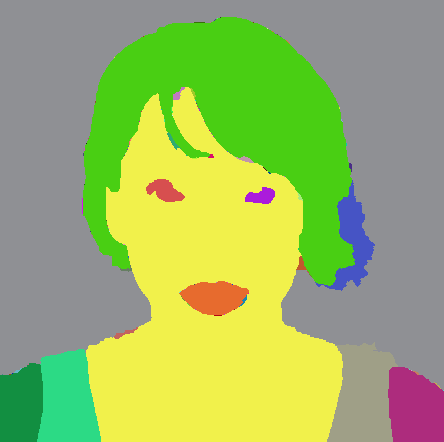}&
\includegraphics[height=0.16\linewidth]{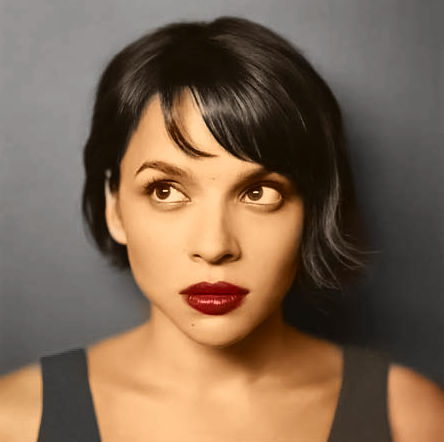}&
\includegraphics[height=0.16\linewidth]{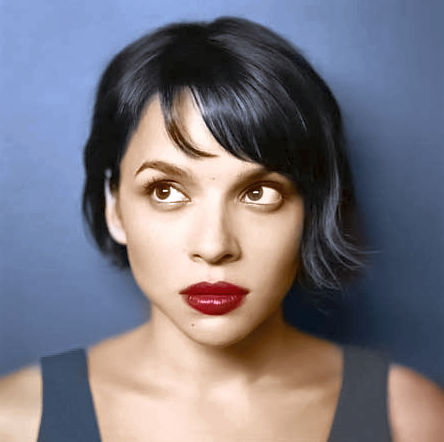}
\\
{\small (a)} & {\small (b)} & {\small (c)} & {\small (d)} & {\small (e)} & {\small (f)}
\end{tabular}
\end{minipage}
 \caption{\label{fig:UserInteraction}
   Result of interactive colorization. (a-c) show the input gray image, automatic segmented image and the colorization result obtained without any user intervention. The colorization result (e) is computed by using the segmented image (d) obtained after merging small image segments with the help of user inputs. (f) shows the final colorization result obtained after performing contrast stretching on the colorized image (e). The reference color images shown in Figure \ref{fig:Portraits} have been used to colorize  input gray image (a).}
\end{figure*}

\begin{figure*}\center
\begin{minipage}[t]{\linewidth}
\centering
\begin{tabular}{c@{\hspace{0.8mm}} c@{\hspace{0.8mm}} c@{\hspace{0.8mm}} c@{\hspace{0.8mm}} |c@{\hspace{0mm}}}
\includegraphics[width=0.19\linewidth]{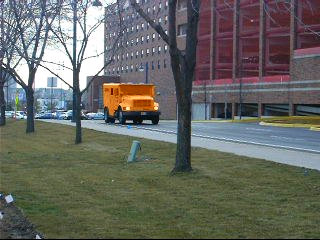}&
\includegraphics[width=0.19\linewidth]{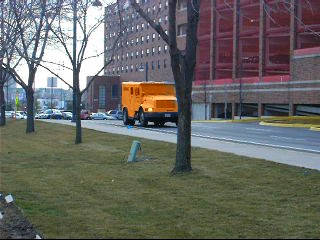}&
\includegraphics[width=0.19\linewidth]{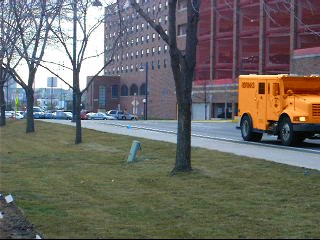}&
\includegraphics[width=0.19\linewidth]{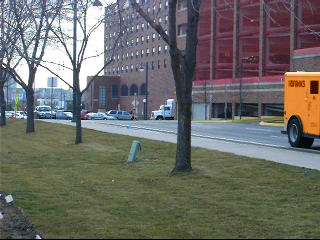}&
\includegraphics[width=0.19\linewidth]{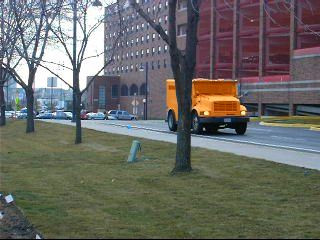}\\
\includegraphics[width=0.19\linewidth]{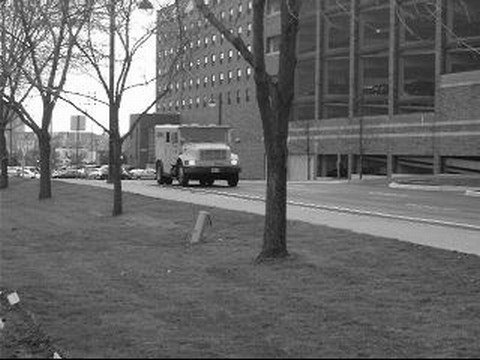}&
\includegraphics[width=0.19\linewidth]{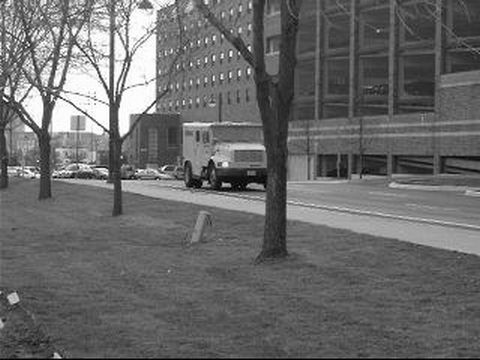}&
\includegraphics[width=0.19\linewidth]{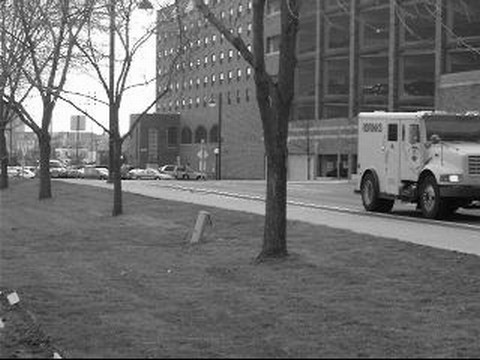}&
\includegraphics[width=0.19\linewidth]{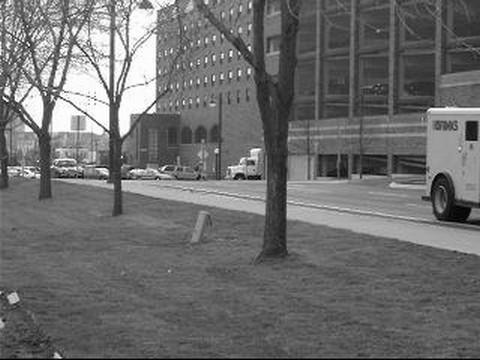}&
\includegraphics[width=0.19\linewidth]{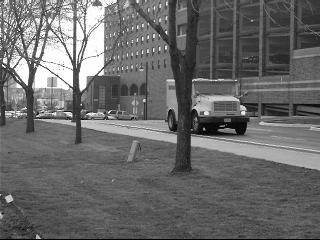}\\
\includegraphics[width=0.19\linewidth]{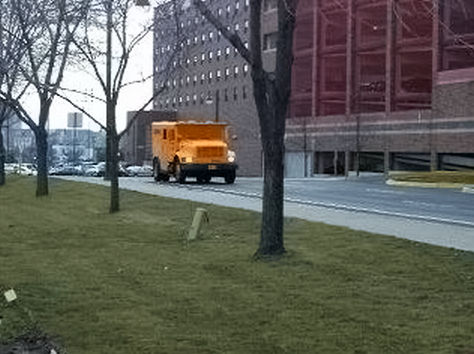}&
\includegraphics[width=0.19\linewidth]{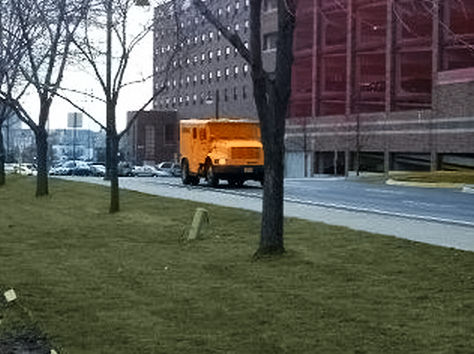}&
\includegraphics[width=0.19\linewidth]{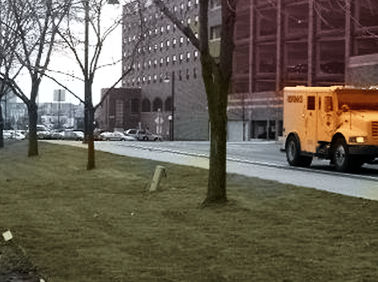}&
\includegraphics[width=0.19\linewidth]{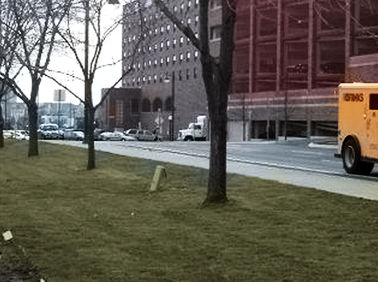}&

\end{tabular}
\end{minipage}
 \caption{Color transfer among video frames. The first row shows the ground truth video frames. The second and third rows show corresponding gray images and colorization results obtained by using the proposed method, respectively. The last column shows the reference color image that has been used by proposed algorithm to colorize all input gray images.}
\label{fig:Video1}
\end{figure*}

\begin{figure*}\center
\begin{minipage}[t]{\linewidth}
\centering
\begin{tabular}{c@{\hspace{0.8mm}} | c@{\hspace{0.8mm}} c@{\hspace{0.8mm}} c@{\hspace{0mm}}}
\includegraphics[width=0.24\linewidth]{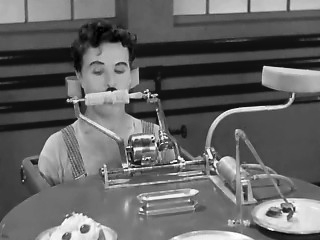}&
\includegraphics[width=0.24\linewidth]{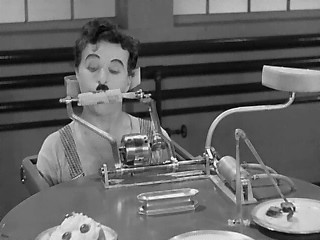}&
\includegraphics[width=0.24\linewidth]{images/Charlie1.jpg}&
\includegraphics[width=0.24\linewidth]{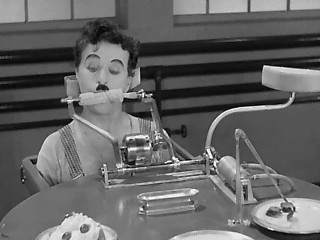}\\
\includegraphics[width=0.24\linewidth]{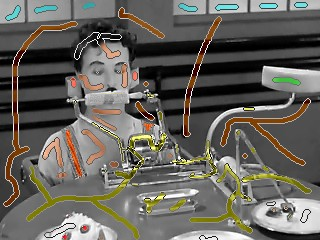}&
\includegraphics[width=0.24\linewidth]{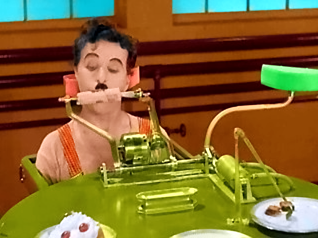}&
\includegraphics[width=0.24\linewidth]{images/Charlie1_result.png}&
\includegraphics[width=0.24\linewidth]{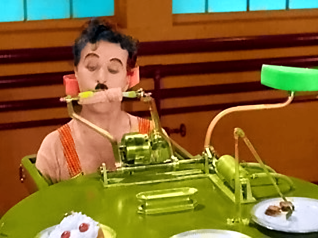}
\end{tabular}
\end{minipage}
 \caption{\label{fig:Charlie}Color transfer among video frames. The first column shows the color scribbled image that has been taken as reference color image. To transfer the color among video frames, first, we propagate the color scribbles across all image pixels by using~\cite{Yatziv06} and then we use this color image as an input to our algorithm to transfer the color information among other video frames. The colorization results are shown in the second, third and fourth columns. \emph{(These video frames have been taken from an old black-and-white Charlie Chaplin's movie Feeding Machine.)} }
\end{figure*}

\section{Experimental Results}
\label{sec:4}
In this section, we have shown the colorization results of the proposed algorithm on a wide variety of images and compared the results with other state-of-the-art colorization algorithms. The colorization results shown in this paper use constant parameter settings that have been reported along with each algorithm step. For a fair comparison, we compared our approach with those colorization algorithms that use only color reference images to transfer the color without any other user inputs.  

Figure \ref{fig:ECCV_SIGGRAPH} compares our colorization results with the algorithms proposed by Welsh $et\ al.$~\cite{Welsh02}, Irony $et\ al.$~\cite{Irony05} and Charpiat $et\ al.$~\cite{Charpiat08}. To compute the colorization results by using other methods, we used the same parameter setting as suggested by their authors. The algorithm~\cite{Irony05} requires the segmentation mask of the reference color image as an additional input. We compute this mask by using automatic image segmentation with same input parameters as we used in Section ~\ref{sec:ColorRefinement}. From the figure, it can be seen that the proposed algorithm yields significantly more visually appealing colorization results as compare to other colorization methods.

Figure \ref{fig:Portraits} shows the colorization results on a set of portrait images. We used multiple reference color images to train the randomized decision forest. These reference images have been shown in the first row. The second row shows the input gray images. The third and fourth rows show the colorization results obtained by another learning-based algorithm~\cite{Charpiat08} and our method, respectively. Figure \ref{fig:Maison} shows the colorization results on another dataset. The input gray images as well as the reference color images have been taken from \emph{Pasadena Houses} Caltech database\footnote{http://www.vision.caltech.edu/html-files/archive.html}. The ground truth images (original color images) and their corresponding gray images are shown in the first and second columns. The third and fourth columns show the colorization results obtained by our method and Charpiat $et\ al.$~\cite{Charpiat08}, respectively. The last column shows the reference color images that have been used by both algorithms to learn the color information. Figure \ref{fig:Maison} shows that our algorithm is able to achieve the colorization results that are very similar to the ground truth images.

Figure \ref{fig:SIG2008} compares the performance of the proposed algorithm with the methods that require additional inputs from the user. The input gray image and color reference image used for comparison, have been adopted from~\cite{Liu08}. It can be seen that the methods~\cite{Liu08} and~\cite{Alex11} are able to retrieve exact color values at some image pixels due to their use of spatial positions during the colorization process. While the use of spatial position helps to achieve better colorization results, it also limits the performance of these algorithms. These algorithms fail in case of the reference image contain different objects or taken from different view angle. Although the proposed algorithm does not use the spatial position, it is able to achieve comparable performance to these algorithms~\cite{Liu08}~\cite{Alex11} while it clearly outperforms the other algorithm~\cite{Irony05}.  

Figure \ref{fig:UserInteraction} shows the result of interactive colorization. The user interaction during the color scribble refinement (Section \ref{sec:ColorRefinement}) by merging the segments that belong to same image region, helps to generate realistic colorization results. Figure \ref{fig:UserInteraction}(b) and \ref{fig:UserInteraction}(d) show the segmentation results obtained before and after the user inputs. The colorization results using these two segmentation masks are shown in figure \ref{fig:UserInteraction}(c) and \ref{fig:UserInteraction}(e), respectively. The quality of colorization results can be further improved by applying image-specific operations such as contrast stretching as shown in figure \ref{fig:UserInteraction}(f).

Figure \ref{fig:Video1} shows the results of video colorization. The first and second rows show the ground truth video fra-mes and corresponding gray images, respectively. The colorization results obtained by using our method have been shown in the third row. The reference color image that has been used to colorize these images has been shown in last column. Traditionally, Optical flow based approaches are used to transfer the color information between video frames. These algorithms often fail in case of large displacement between video frames. However, as we discussed earlier the proposed algorithm does not rely on spatial positions, it transfers the color between video frames successfully even in case of large displacement as shown in the figure.

Figure \ref{fig:Charlie} shows the colorization results on three consecutive video frames that have been taken from an old black-and-white Charlie Chaplin's movie \emph{Feeding Machine}. Here, we take a color scribbled frame as a reference color image shown in first column. To transfer the color between video frames, first, we propagate the color scribbles across all image pixels by using~\cite{Yatziv06} and then we use this color image as an input to our algorithm to transfer the color information among other video frames. The colorization results are shown in the second, third and fourth columns.
\subsection{Limitations}
There are few limitations in our method. First, our use of superpixel representation, while supporting more spatial coherency in colorization, can be inaccurate at object boundaries or thin image structures. This could potentially lead to bleeding artifacts at the boundaries. Second, in Section~\ref{sec:ColorRefinement},  we perform the image space voting on image segments generated by using fixed parameter settings. While such parameters settings work well for few images, it generates very small segments in dense textured image regions. During the color scribble refinement step, we skip all such image segments due to very few superpixels within it for voting (requires at least three superpixels). These image segments sometime lead to some artifacts in colorization results. By merging these small image segments with the help of user inputs or by adjusting the input parameters of image segments, the quality of the colorization can be further improved.  In video colorization, we process each frame independently which may lead to some temporal artifacts (e.g. random color changes) in non-textured image regions of the resulting colorized video.

\section{Conclusion}
\label{sec:5}
In this paper, we proposed a learning-based approach to colorize input gray image by using one or more color reference images without any user intervention. The comparison results with other state-of-the-art colorization algorithms show that the proposed algorithm works well on a diverse variety of images and outperforms existing colorization approaches. We have shown that the proposed algorithm can also be used to transfer the color information between the video frames. The algorithm performs well even in case of occlusion or large displacement in object's position between the video frames. We have also shown that little user interaction during color scribble refinement step improves the results significantly and able to achieve realistic colorizations.

\end{document}